\documentclass[12pt]{iopart}
\usepackage{bm}
\usepackage{graphicx}
\usepackage{color}    
\usepackage[english]{babel}
\usepackage{mathtools}
\usepackage{latexsym}
\usepackage{hyperref}
\usepackage{xspace}
\usepackage{amssymb}
\usepackage{amsthm}
\usepackage{perpage} 

\MakePerPage{footnote}

\newcommand{\ket}[1]{\left|{#1}\right\rangle}
\newcommand{\bra}[1]{\left\langle{#1}\right|}

\newcommand{\beq}{\begin{equation}}
\newcommand{\eeq}{\end{equation}}

\begin{document}
\title{Experimental Verification of Quantum Discord in Continuous-Variable States}
\author{S. Hosseini$^1$\footnote{Corresponding 
author: sara.hosseini@anu.edu.au}
, S. Rahimi-Keshari$^2$\footnote{Corresponding 
author: s.rahimik@gmail.com}, J.Y. Haw$^1$, S. M. Assad$^1$, H. M. Chrzanowski$^1$, J. Janousek$^1$, T. Symul$^1$, T. C. Ralph$^2$, P. K. Lam$^1$}

\address{$^1$Center for Quantum Computation and Communication Technology, Department of Quantum Science, The Australian National University, Canberra, ACT 0200, Australia 
\\$^2$ Center for Quantum Computation and Communication Technology, School of Mathematics and Physics, University of Queensland, St Lucia, Queensland 4072, Australia}
\date{\today}

\begin{abstract}
 We introduce a simple and efficient technique to verify quantum discord in unknown Gaussian states and a certain class of non-Gaussian states. We show that any separation in the peaks of the marginal distributions of one subsystem conditioned on two different outcomes of homodyne measurements performed on the other subsystem indicates correlation between the corresponding quadratures, and hence nonzero discord.
We also apply this method to non-Gaussian states that are prepared by overlapping a statistical mixture of coherent and vacuum states on a beam splitter. We experimentally demonstrate this technique by verifying nonzero quantum discord in a bipartite Gaussian and certain class of non-Gaussian states. 

\end{abstract}

\maketitle

\section{Introduction}
Quantum correlations have been the subject of many studies during the last decades, in particular, as a resource for quantum information processing and quantum communication. Previously, any correlation in the absence of entanglement was thought to be purely classical as they can be prepared with local operations and classical communications. However, there are reasons to believe that this was not the whole story; for example, there are quantum computational models with no or little entanglement, which can efficiently perform tasks that are believed to be classically hard~\cite{KL98,DFCaves05}. Quantum discord was introduced as a general measure of quantum correlation that can capture nonclassical correlations beyond entanglement~\cite{Zurek01}. Discord was suggested as a figure of merit for characterizing the quantum resources in a computational model~\cite{DSCaves08}; it also was introduced as a resource for quantum state merging~\cite{madhok11,cavalcanti} and for encoding information onto a quantum state~\cite{gu}. This measure of nonclassical correlation has been extended to continuous-variable systems to study quantum correlations in Gaussian states~\cite{Datta10,Paris10} and certain non-Gaussian states~\cite{TMAdessoK12}.

Considering the importance of quantum discord, of particular interest is to experimentally verify discord for an \textit{unknown} quantum system. Methods have been proposed to test for nonvanishing quantum discord of bipartite discrete-variable quantum states~\cite{rahimi2010,BC10,DVBrukner10,modi2011,ZYCO11,GAdesso12,serra2012}, some of which have been experimentally implemented in nuclear-magnetic-resonance systems~\cite{Serra11,Laflamme11} and in an optical system~\cite{walborn2012}.
Recently a measurement-based method for verifying quantum discord was introduced~\cite{rahimi-keshari2013}, which can be applied to both discrete- and continuous-variable systems. 

Here we introduce and demonstrate a simple and efficient experimental technique for verifying quantum discord in Gaussian states. It was shown that the ``if and only if'' condition for a bipartite Gaussian state to have zero discord is that there is no correlation between the quadratures of two subsystems, i.e., it is a product state~\cite{rahimi-keshari2013}. In our method, we use two homodyne detections to examine the correlations between quadratures of subsystems $A$ and $B$. 
For example, if the peaks of the conditional marginal distributions of $B$'s quadrature corresponding to the positive and negative outcomes of homodyne measurements performed on $A$'s quadrature, do not coincide at the same point, those quadratures are correlated.
In order to consider all possible correlations, we check the correlations between all four combinations of the amplitude and phase quadratures of A and B. If at least one of them is found to be correlated, quantum discord is nonzero, otherwise it is zero. There is also a simple way to verify quantum discord in bipartite non-Gaussian states prepared by subjecting a statistical mixture of coherent states to one port of a beam splitter while the other port is in the vacuum state. We show that any changes in the conditional marginal distributions observed using our method for this class of bipartite non-Gaussian states indicate nonzero discord. We experimentally demonstrate our technique by preparing Gaussian and non-Gaussian states with no entanglement and verify the presence of quantum discord.\\
\textcolor{white}{a}\hspace{1em}This paper is structured as follows. In Section \ref{sec:theory}, we review the theoretical description of quantum discord and introduce our technique to experimentally verify quantum discord in Gaussian states and certain class of non-Gaussian states. In Section \ref{sec:experiment}, we thoroughly describe the experiments which are performed to examine this method on a Gaussian state and three different non-Gaussian states, and the experimental results are presented in detail. Finally, Section \ref{conclude} concludes our main findings.

\section{Theory}
\label{sec:theory}
\subsection{Quantum Discord}
\label{subsec:QD}
 Quantum discord is defined as the mismatch between two quantum analogues of classically equivalent expressions of the mutual information~\cite{Zurek01}. For two classical random variables $A$ and $B$, the total correlation is given by mutual information, which can be defined by two equivalent expressions $I(A:B){=}H(A){+}H(B){-}H(A,B)$ and $J(A:B){=}H(A){-}H(A|B){\equiv }H(B){-}H(B|A)$, where $H(X)$ is the Shannon entropy and $H(X|Y)$ is the conditional entropy.
For a bipartite quantum system, the quantum mutual information is defined by $I(\rho_{AB}){=}S(\rho_A){+}S(\rho_B){-}S(\rho_{AB})$ that is analogous to $I(A:B)$, where $S(\rho){=}-\text{Tr}[\rho \log_2(\rho)]$ is the von Neumann entropy. 
A measurement-based quantum conditional entropy is $S_{\{\Pi_j\}}(A|B){=}\sum_{j}p_{j}S(\rho_{A|j})$, where  $p_j{=}\text{Tr}[\rho_{AB}\Pi_{j}]$ is the probability of obtaining the conditional state $\rho_{A|j}{=}\text{Tr}_B[\rho_{AB}\Pi_j]/p_j$, and the set $\{\Pi_j\}$, with $\sum_j\Pi_j{=}\mathbb{I}$, form a positive operator-valued measurement (POVM) performed on subsystem $B$.
As this quantity is measurement dependent, the quantum version of the expression including conditional entropy is defined as $J^{\leftarrow}(\rho_{AB}){=}S(\rho_{A}){-}\text{min}_{\{\Pi_j\}}S_{\{\Pi_j\}}(A|B)$, which is known as one way classical correlation.The minimization is performed over all possible measurements.
Therefore, the quantum discord from $B$ to $A$ is defined as:
\begin{align}
D^{\leftarrow}(\rho_{AB})&=I(\rho_{AB})-J^{\leftarrow}(\rho_{AB}) \nonumber \\
&=S(\rho_{B})-S(\rho_{AB})+\text{min}_{\{\Pi_j\}}S_{\{\Pi_j\}}(A|B)\;.
\end{align}
 
In general, it is not clear how to perform the minimization for any arbitrary state, unless there are restrictions to certain class of states and POVMs. Gaussian quantum discord is defined as the quantum discord of a bipartite Gaussian state, where the minimization is restricted to generalized Gaussian measurements~\cite{Datta10,Paris10}. This quantity was experimentally estimated and characterized for a two-mode squeezed thermal state~\cite{blandino2012}, two-mode squeezed vacuum state generated by a four-wave mixing process~\cite{Vogl2013}, and entangled and separable Gaussian states~\cite{Madsen2012}. Gaussian states with nonzero discord are shown to be used to reveal interference~\cite{Meda2013}. It was recently shown that Gaussian states with nonzero Gaussian discord have nonzero discord~\cite{rahimi-keshari2013}.

\subsection{Verification of Quantum Discord in Gaussian States}
\label{subsec:theoryG}
The measurement-based method for verifying quantum discord~\cite{rahimi-keshari2013} is based on measuring the conditional states of subsystem $B$ corresponding to the outcomes of an informationally complete POVM~\cite{Prug77,Busch91} performed on subsystem $A$. If the conditional states commute with one another then quantum discord is zero, otherwise is nonzero. However, if some prior knowledge about the state is available, one may be able to verify discord with only a few measurements. It was shown in ~\cite{rahimi-keshari2013} that in principle for Gaussian states nonvanishing quantum discord can be verified by checking whether the peaks of two conditional Wigner functions corresponding to two different outcomes of heterodyne measurements do not coincide at the same point in the phase space. However, in practice, this is not efficient, as one has to repeat the measurements many times in order to obtain sufficient data for finding the peaks of the conditional Wigner functions.  
Here we introduce a simple and efficient experimental technique for verifying discord of Gaussian states, which can be also applied to some class of non-Gaussian states.
 
In general, one can always characterize Gaussian states in terms of the means and covariance matrix of their quadratures $x$ and $p$~\cite{Adesso-Illuminati}.
For a bipartite system with modal annihilation operators $\hat a{=}\hat{x}_A{+}i\hat{p}_A$ and $\hat b{=}\hat{x}_B{+}i\hat{p}_B$, we define quadrature vectors for each subsystem, $\hat{\mathbf{x}}_A{=}(\hat{x}_A,\hat{p}_A)$ and $\hat{\mathbf{x}}_B{=}(\hat{x}_B,\hat{p}_B)$, and an overall quadrature vector $\hat{\mathbf{x}}{=}(\hat{\mathbf{x}}_A,\hat{\mathbf{x}}_B){=}(\hat{x}_A,\hat{p}_A,\hat{x}_B,\hat{p}_B)$.
The vector $\bar{\mathbf{x}}$ represents the means of the quadratures, $\bar{\mathbf{x}}=\langle\hat{\mathbf{x}}\rangle $, and the covariance matrix is 
\begin{equation}
\bm{\sigma}=\langle|\hat{\mathbf{x}}^T\hat{\mathbf{x}}|\rangle-\bar{\mathbf{x}}^T\bar{\mathbf{x}}=
\begin{pmatrix}
\mathbf{A} & \mathbf{C} \\
\mathbf{C}^{T} & \mathbf{B}
\end{pmatrix}
\;,
\label{sigmagen}
\end{equation}
where we define $|\hat{\mathbf{x}}_i\hat{\mathbf{x}}_j|=\frac{1}{2}(\hat{\mathbf{x}}_i\hat{\mathbf{x}}_j+\hat{\mathbf{x}}_j\hat{\mathbf{x}}_i)$, and $\mathbf{A}$, $\mathbf{B}$ and $\mathbf{C}$ are 2$\times$2 matrices. The Wigner function is then given by
\begin{equation}
W_{AB}(\mathbf{x})=\frac{1}{4\pi^2\sqrt{\det\bm{\sigma}}}
\label{wigner}\,\exp\!\left(-\frac{(\mathbf{x}-\bar{\mathbf{x}})\bm{\sigma}^{-1}(\mathbf{x}-\bar{\mathbf{x}})^T}{2}\right),
\end{equation}
A bipartite Gaussian state has zero discord if and only if there is no correlation between the quadratures of the two subsystems, i.e., 
\cite{rahimi-keshari2013}
\begin{equation}
\mathbf{C}=
\begin{pmatrix}
c_{11} & c_{12} \\
c_{21} & c_{22}
\end{pmatrix}
=0\;.
\end{equation}
 
Suppose Alice and Bob are sharing a bipartite Gaussian state. In order to verify quantum discord they use two homodyne detections, one for each subsystem. Without loss of generality, we assume $\mathbf{A}{=}\text{diag}(a1,a2)$, $\mathbf{B}{=}\text{diag}(b1,b2)$ and $\bar{\mathbf{x}}=0$, as these can be always accomplished by appropriately choosing the zero reference phase of the local oscillators and shifting the zero reference points of the quadratures being measured. The joint marginal distribution describing the outcomes of two homodyne detections is then given by~\cite{U97}
\begin{align}
D_{AB}(x_A,\theta_A &,x_B,\theta_B)=\int_{-\infty}^{+\infty}\int_{-\infty}^{+\infty}
dp_A dp_B W(\mathbf{x}\mathbf{U}_{\theta_A,\theta_B})\nonumber \\
&=\frac{\pi}{\sqrt{\lambda_{\theta_{A}}\mu_{\theta_{B}}-\nu_{\theta_{A},\theta_{B}}^2}} \nonumber \\
&\times \exp\left(-\lambda_{\theta_{A}} x_A^2-\mu_{\theta_{B}} x_B^2+2\nu_{\theta_{A},\theta_{B}} x_A x_B\right),
\end{align}
where 
\begin{equation}
\mathbf{U}_{\theta_A,\theta_B}=
\begin{pmatrix}
\cos\theta_A & \sin\theta_A &0&0\\
-\sin\theta_A & \cos\theta_A&0&0\\
0&0&\cos\theta_B & \sin\theta_B\\
0&0&-\sin\theta_B & \cos\theta_B \nonumber
\end{pmatrix}
\end{equation}
with $\theta_A$ and $\theta_B$ being the phases of the local oscillators used in Alice's and Bob's homodyne detection, respectively, and $\lambda_{\theta_{A}}$, $\mu_{\theta_{B}}$, and $\nu_{\theta_{A},\theta_{B}}$ are some functions of the  covariance matrix elements, which depend on $\theta_A$ and $\theta_B$. If $\nu_{\theta_{A},\theta_{B}}$ is nonzero, then the quadrature associated with the phase $\theta_{A}$ of subsystem $A$ is correlated to the quadrature associated with the phase $\theta_{B}$ of subsystem $B$. In order to check this, Bob measures two conditional marginal distributions corresponding to outcomes $x_A>0$ and $x_A<0$ of Alice's measurements
\begin{align}
\label{eq:condtion}
D_{B|\pm }(x_B,\theta_B,\theta_A)&=\int_{0}^{\pm \infty}(\pm 1) dx_A D_{AB}(x_A,\theta_A ,x_B,\theta_B) \nonumber \\
&=\frac{\sqrt{\pi\lambda_{\theta_{A}}}\exp\left({\frac{\nu_{\theta_{A},\theta_{B}}^2-\mu_{\theta_{B}}\lambda_{\theta_{A}}}{\lambda_{\theta_{A}}}x_B^2}\right)}{\sqrt{\mu_{\theta_{B}}\lambda_{\theta_{A}}-\nu_{\theta_{A},\theta_{B}}^2}} \nonumber\\
&\times\left(1\pm \text{Erf}\left(\frac{\nu_{\theta_{A},\theta_{B}} x_B}{\sqrt{\lambda_{\theta_{A}}}}\right)\right),
\end{align}
where $\text{Erf}(.)$ being the error function. If the peaks of the marginal distributions $D_{B|+ }(x_B,\theta_B,\theta_A)$ and $D_{B|- }(x_B,\theta_B,\theta_A)$ do not coincide with one another, this implies that $\nu_{\theta_{A},\theta_{B}}\neq0$.
 
Using this technique Alice and Bob can now verify quantum discord. As we have
\begin{align}
\nu_{0,0}&=\frac{c_1}{2a_1b_1-2c_1^2} , \nonumber\\
\nu_{0,\frac{\pi}{2}}&=\frac{c_2}{2a_1b_2-2c_2^2},\nonumber\\
\nu_{\frac{\pi}{2},0}&=\frac{c_3}{2a_2b_1-2c_3^2},\nonumber\\
\nu_{\frac{\pi}{2},\frac{\pi}{2}}&=\frac{c_4}{2a_2b_2-2c_4^2},\nonumber
\end{align}
they only need to choose the phases of their local oscillator to be 0 or $\pi/2$ and measure the conditional marginal distribution $D_{B|\pm }(x_B,\theta_B,\theta_A)$ to check whether the elements of matrix $\mathbf{C}$ are zero or not. If at least one of the elements is found to be nonzero, the state has nonzero quantum discord.  
 
\subsection{Verification of Quantum Discord in Non-Gaussian States}
\label{subsec:theoryNG}
One way to create quantum states with nonclassical correlation is to use beam splitter. It was shown that nonclassicality of input states to a beam splitter is a necessary condition for generating entanglement at the output of a beam splitter~\cite{kim2002, wang2002}. Here we show that bipartite quantum states that are prepared by subjecting a statistical mixture of coherent states to a beam splitter, while the other port is in the vacuum state, have nonzero discord. We show that quantum discord for this class of non-Gaussian states can be simply verified.

By using the Glauber-Sudarshan representation~\cite{Glauber,Sudarshan} for an input state to a beam splitter
\begin{equation}
\rho_1\otimes\ket{0}\bra{0}=\int d^2\alpha P_1(\alpha) \ket{\alpha}\bra{\alpha}\otimes \ket{0}\bra{0},
\end{equation}    
the output state is then given by
\begin{equation}
\rho_{\text{out}}=\int d^2\alpha P_1(\alpha) \ket{\eta\alpha}\bra{\eta\alpha}\otimes \ket{\tilde\eta \alpha}\bra{\tilde\eta \alpha},
\label{rho-out}
\end{equation}  
where $\eta$ is the transmissivity of the beam splitter and $\tilde\eta=\sqrt{1-\eta^2}$. If $P_1(\alpha)$ is a positive semidefinite Gaussian or non-Gaussian function other than the Dirac delta function, the state $\rho_{\text{out}}$ has nonzero discord, as it is a mixture of nonorthogonal states of two subsystems \cite {rahimi-keshari2013}. 
 
The Wigner function of the state after the beam splitter is given by~\cite{U97}
\begin{align}
W_{\text{out}}(x_1,&p_1,x_2,p_2)=W_1(\eta x_1+\tilde\eta x_2,\eta p_1+\tilde\eta p_2)\nonumber \\
&\times\frac{1}{\pi}\exp\left[-(\eta x_2-\tilde\eta x_1)^2-(\eta p_2-\tilde\eta p_1)^2 \right].
\end{align}
where $W_1(x,p)$ is the Wigner function for the input state $\rho_1$. Therefore, the necessary and sufficient condition to verify discord in the state \eqref{rho-out} is to check whether the Wigner function of any of marginal states at the output, for example
\begin{equation}
W_{\text{out},1}(x_1,p_1)=\int_{-\infty}^{+\infty}\int_{-\infty}^{+\infty} dx_2 dp_2 W_{\text{out}}(x_1,p_1,x_2,p_2)\nonumber ,
\end{equation}  
is the Wigner function of a coherent state or not.
 
Also by applying our technique developed in the previous subsection, if one observes any changes in the conditional marginal distributions, that indicates correlation between the two quadratures and hence nonzero quantum discord. By measuring $x$-quadratures of two subsystems using two homodyne detections, the joint marginal distribution is then given by
\begin{equation} 
D(x_1,x_2)= \frac{1}{\sqrt{\pi}} D_1(\eta x_1+\tilde\eta x_2) e^{-(\eta x_2-\tilde\eta x_1)^2},
\end{equation}
where $D_1(x)$ is the marginal distribution of $W_1(x,p)$. If the input state is not a coherent state then $\rho_{out}$ has discord, otherwise zero discord.
In the following section, we demonstrate the use of our technique for three different non-Gaussian states.     
 
Notice that our technique has limited use in verifying quantum discord of completely general non-Gaussian states where any peak separation is not necessarily an indication of quantum discord. For example, this state  
\begin{align}
\rho_{AB}=&\frac{1}{4}\big(\ket{\alpha}\bra{\alpha} \otimes (\ket{0}+\ket{1})(\bra{0}+\bra{1}) \nonumber \\              
&+ \ket{-\alpha}\bra{-\alpha} \otimes (\ket{0}-\ket{1})(\bra{0}-\bra{1})\big) \nonumber
\end{align}
has zero discord from $B$ to $A$, but by using our method one can see that there is a peak separation in the conditional marginal distributions of $B$ .
There are also quantum states with nonzero discord from $B$ to $A$ but no peak separation in the conditional marginal distributions; one such state 
is
\begin{equation}
\rho_{AB}=\rho_{A,1}\otimes\rho_{B,th}+\rho_{A,2}\otimes\rho_{B,S},
\end{equation}
where $\rho_{B,th}$ and $\rho_{B,S}$ are two non-commuting states, thermal state and squeezed vacuum state, respectively, and $\rho_1$ and $\rho_2$ are two arbitrary states.

\section{Experiment}
\label{sec:experiment}
\subsection{Quantum Discord in Gaussian States}
\label{sec:expG}

\begin{figure*}
\centering
\includegraphics[width=12cm]{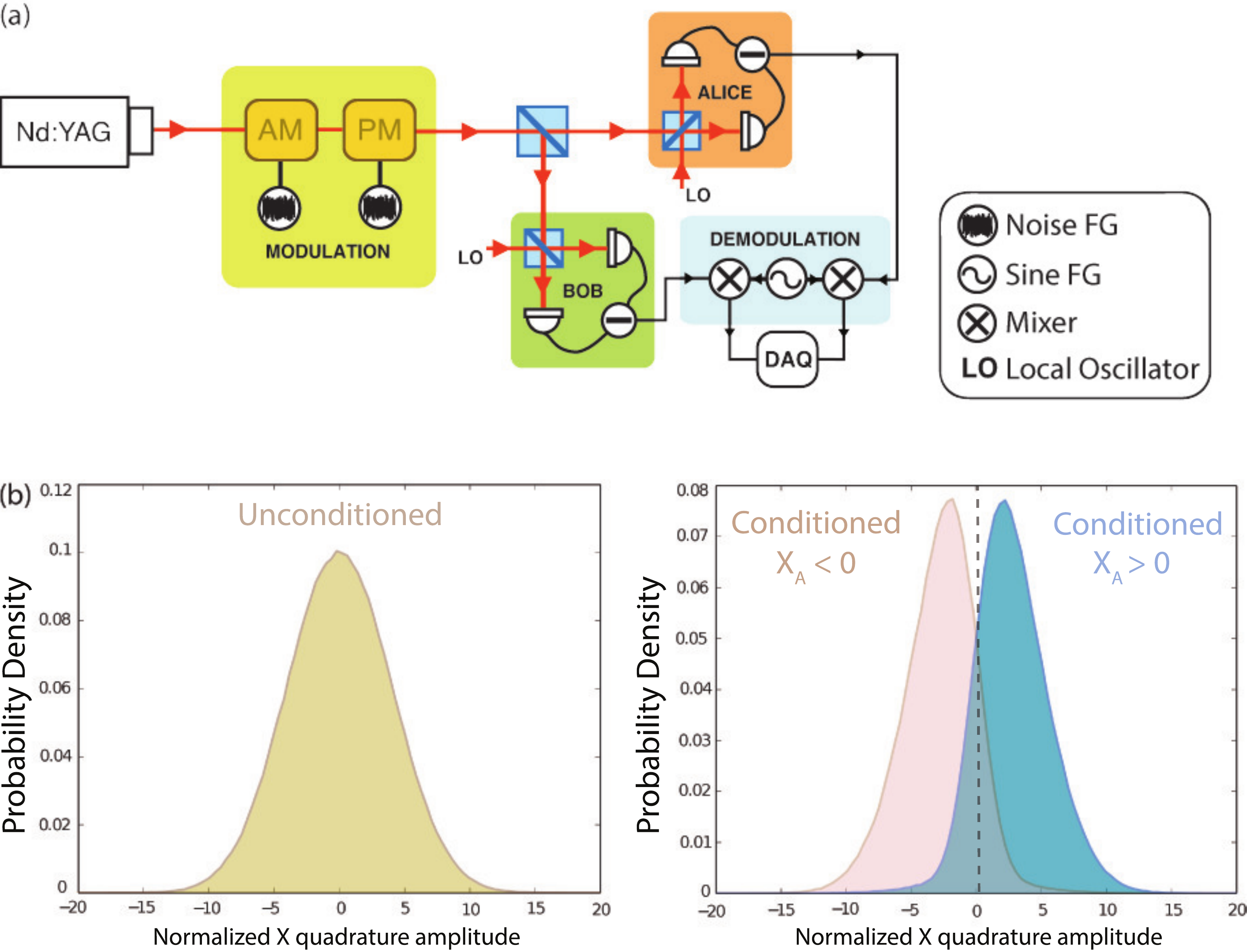} 
 \caption{ (a) Schematic diagram of the experimental setup. Here, AM and PM are the electro-optic modulators (EOM) driven by function generators (FG), which in turn provide displacement of the vacuum state in amplitude and phase quadrature with Gaussian distributed noise. Laser light is passed through EOMs and is split on 50:50 beam splitter. Each part is sent to a homodyne measurement station (Alice and Bob). Collected data points from each homodyne station are demodulated and sampled using a digital data acquisition system (DAQ). (b) The unconditioned (left) and conditioned (right) probability distributions of the bipartite Gaussian state with discord.  The state is obtained from a Gaussian distributed modulated beam with modulation depth of 4.5 times the quantum noise. The blue and pink shaded curves show the probability distributions conditioned respectively on $x_A>0$ and $x_A<0$, where $x_A$ is the measured amplitude quadrature of subsystem $A$ normalized to quantum noise. The peak separation indicates that the states A and B are discordant.}

   \label{fig:schematic}
 \end{figure*}
The experimental setup used to verify the presence of quantum discord is depicted in Figure \ref {fig:schematic} (a). The laser light is passed through a mode cleaner cavity to provide a quantum noise limited light source. A large portion of it, is used as the bright source of local oscillator for homodyne detection, and a small portion, is passed through a pair of phase and amplitude elctro-optic modulators (EOMs). EOMs are used to provide Gaussian distributed modulation on both quadratures. The modulated beam is then split on a 50:50 beam splitter to generate two separable but correlated bipartite state (A and B). Each part of it, is sent to a homodyne measurement station, which we labelled Alice and Bob.

Following subsection \ref{subsec:theoryG}, in order to check whether the elements of matrix $\mathbf{C}$ are zero or not, all possible correlations between two subsystems $A$ and $B$ need to be checked. In order to do that we first lock Bob's station to amplitude quadrature and perform homodyne measurements on both of the stations by locking Alice's station to amplitude quadrature, followed by phase quadrature. The same procedure is repeated for phase quadrature of Bob's station. The marginal distributions of Bob's state conditioned on  Alice's outcomes, $x_A>0$ and $x_A<0$ , are calculated and any possible separation between the peaks of conditional marginal distributions are investigated. In our experiment, the bipartite Gaussian state have correlations in both phase and amplitude quadratures but with very little cross-correlation between the quadratures of two subsystems (See  \ref{appendix}). Hence when Alice and Bob are both locked to the same quadrature, we observe separation between peaks of conditional marginal distributions, as shown in Figure \ref {fig:schematic}(b) for amplitude quadrature.  Similar result is obtained when both subsystems are locked to phase quadrature. As discussed in subsection \ref{subsec:theoryG}, for Gaussian state the peak separation in the conditional marginal distributions is a necessary and sufficient condition of non-zero quantum discord. Hence from our result we conclude that we have a discordant bipartite Gaussian state. The covariance matrix of this bipartite state is presented in  \ref{appendix}, showing  $\mathbf{C}$ is indeed non-zero.
  
In our experiment each pair of detectors are balanced electronically, providing 30 dB of common mode rejection. Typical suppression of cross correlation between orthogonal quadrature is around 25 dB. For each separate homodyne detection, $2.4\times10^6$ data points are sampled at $14\times10^6$ samples per second utilizing a digital acquisition system. In order to provide adequate statistics, this procedure is taken over five times for each data point. These data are then down sampled and digitally filtered to 2-5 MHz. Our homodyne efficiency is typically $96.6\%$, with fringe visibility of  $97.6\%$, generally limited by the mode distortions introduced by the EOMs and the photodiode quantum efficiency of $99\%$.
 
We also investigate the effect of variation of modulation depth on the peaks separation of conditional marginal distributions. This is done by changing the variance of Gaussian noise introduced by (EOM) on the desired quadrature. Since we only modulate the phase quadrature, both subsystems are locked to this quadrature. We apply $22$ different modulation depths on the phase quadrature, ranging from zero to 5 times the quantum noise. For each homodyne detection, $1.2\times10^5$ data points are sampled at 200 ksamp per second and then down sampled at 4 MHz sideband. The process is repeated 20 times 
in order to provide sufficient statistics. 
\begin{figure}[t]
\centering
\includegraphics[width=12cm]{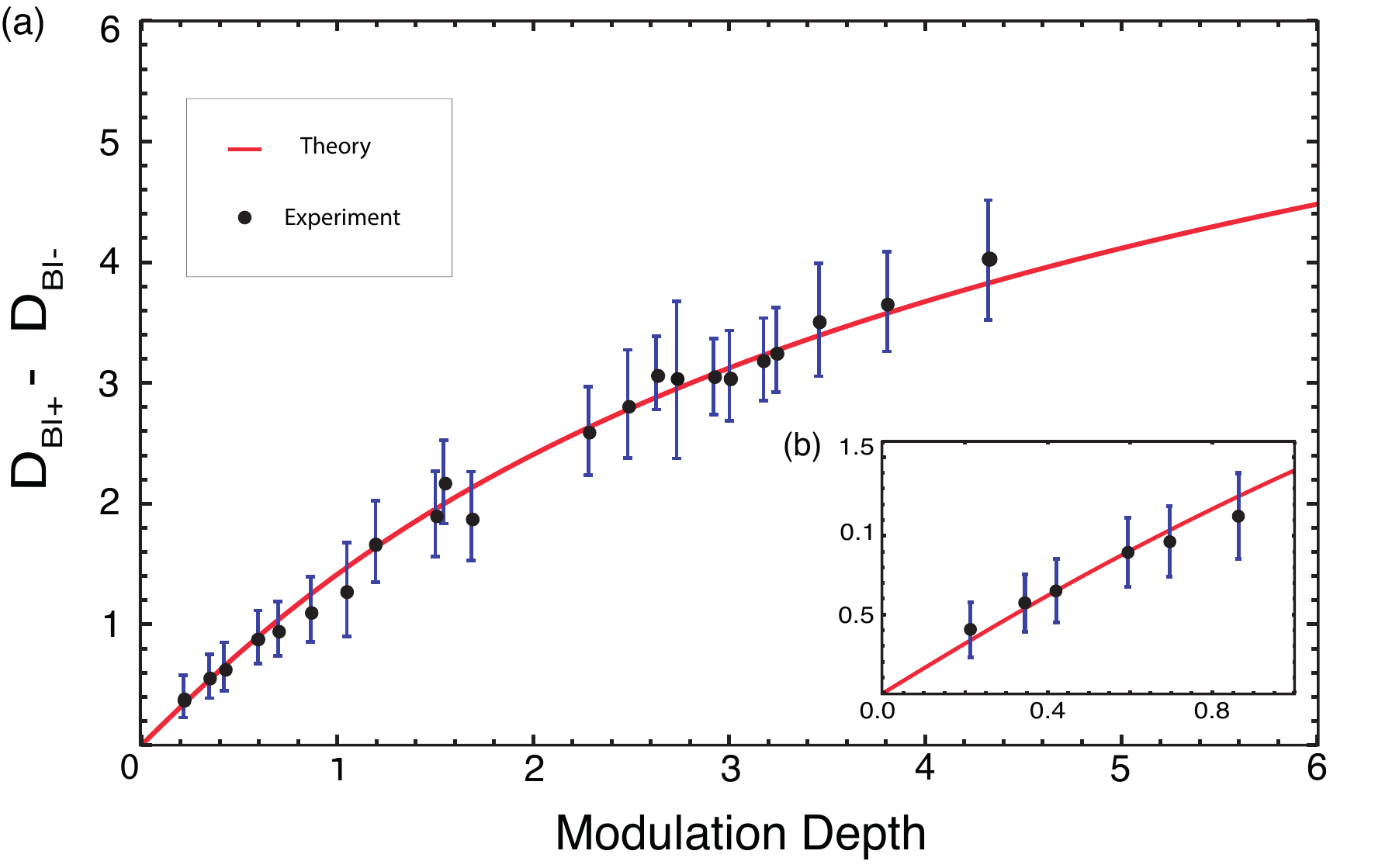}
\caption{(a) Variation of peak separations of marginal distributions conditioned on two different homodyne outcomes, $D_{B|+}-D_{B|-}$, versus modulation depth. The theoretical curve is evaluated according to Eq. \eqref{eq:condtion}. The experimental error bars are estimated using statistical uncertainties. Inset (b) shows the zoom-in for small modulation depth. Even for the smallest modulation depth (0.2 times of quantum noise), our technique is still able to reveal the presence of quantum discord.}
\label{fig:modvar}
\end{figure} 
For each modulation depth, the conditional marginal distributions are evaluated and the separation between two peaks is measured. As shown in Figure \ref{fig:modvar}(a), the separation of the peaks increases monotonically with the modulation depth. This is consistent with the theoretical curve plotted according to Eq. \eqref{eq:condtion}. As the modulation depth increases, more noise is applied on the input beam and thus increases the variance of the input beam. This gives rise to output beams with higher correlations, and hence larger elements of matrix $\mathbf{C}$. It is remarkable that despite the simplicity of our technique, it is robust enough to verify the presence of discord in weakly correlated bipartite Gaussian states, as indicated in the Fig \ref{fig:modvar}(b).

 \subsection{Quantum Discord in Non-Gaussian States}
 \begin{figure*}[t]
 \centering
\includegraphics[width=12cm]{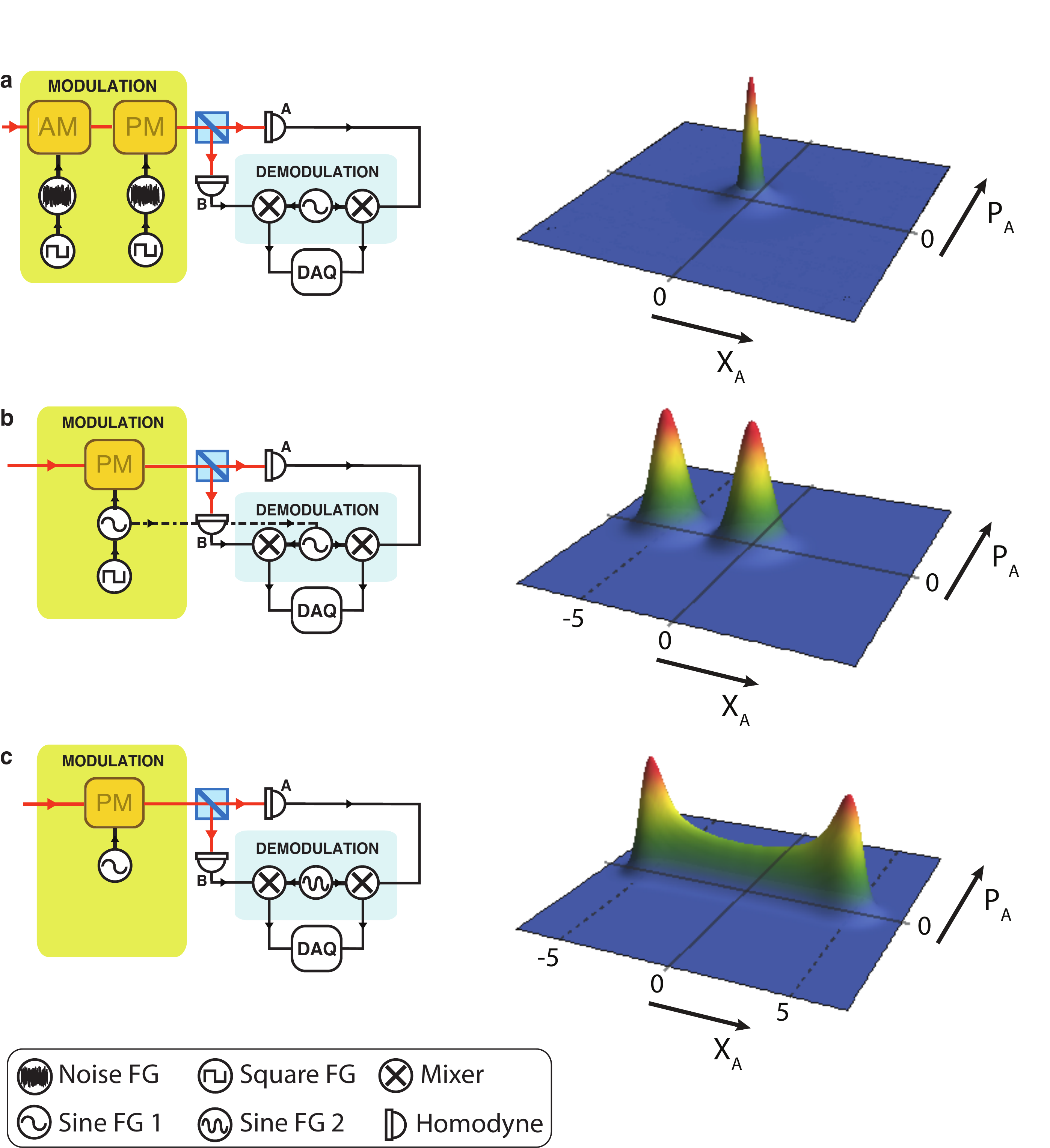} 
 \caption{ Schematic diagram of the modulation and demodulation arrangements used in preparation of the non-Gaussian states (left) and their corresponding positive-definite non-Gaussian Wigner functions (right), $X_{A}$ and $P_{A} $ are normalized quadrature amplitudes (a) Switched noise modulation: This vacuum-thermal superposition state is generated by gating Gaussian noise modulation on both quadratures with square waves; (b) Switched phase modulation: This state is an equal statistical mixture of a vacuum and a coherent state, created by gating a sine wave modulation with a low frequency square wave; (c) Asynchronous detection: This state is prepared by modulating one quadrature with sine wave and demodulating it with another sine wave of slightly different frequency.}
 \label{fig:ngschematic}
 \end{figure*}

 As discussed in subsection \ref{subsec:theoryNG}, our discord verification technique can be applied to bipartite non-Gaussian states obtained by overlapping a statistical mixture of coherent states and vacuum state on a beam splitter. It was previously reported in ~\cite{Jinwei Wu} that a mixture of coherent states can be generated by subjecting a laser beam to time varying modulation. Here, we demonstrate our verification technique to examine quantum discord in non-Gaussian states discussed in Section \ref{subsec:theoryNG}.   In the following, we describe the preparation of three non-Gaussian states with positive-definite  Wigner functions (see Figure \ref{fig:ngschematic}) and discuss the corresponding verification results.

\begin{figure*}[t!]

\centering
\includegraphics[width=12cm]{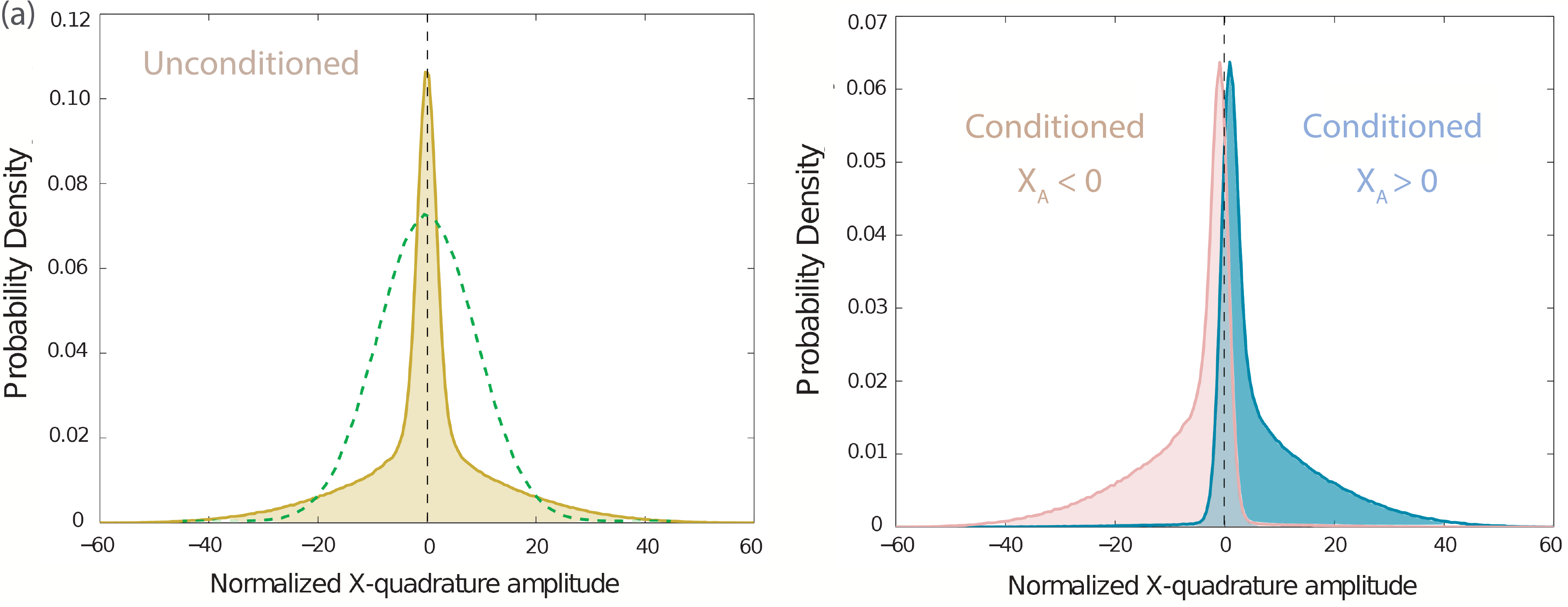}
\includegraphics[width=12cm]{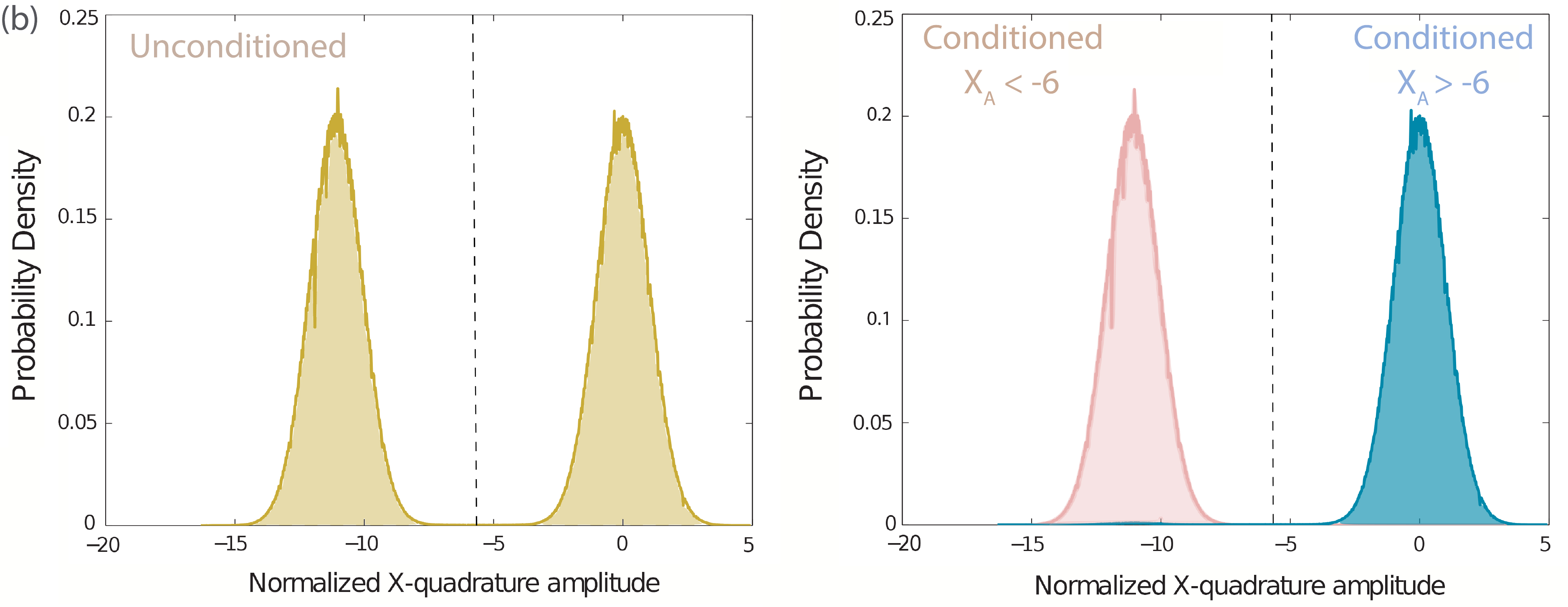}
\includegraphics[width=12cm]{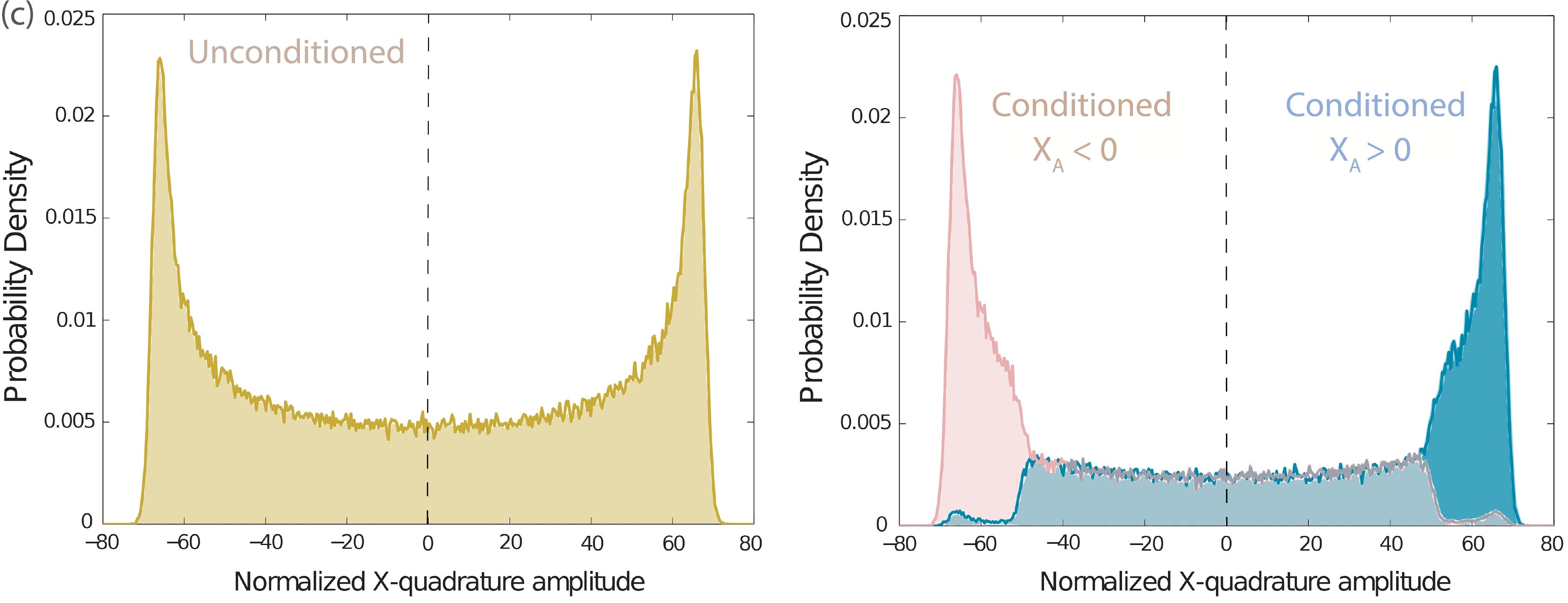}
\caption{ Unconditional (Brown) and conditional probability distributions of two different outcomes (Pink and Blue) of the non-Gaussian states prepared by (a) Switched Noise Modulation (The green dashed curve corresponds to a Gaussian state with average variance of the two Gaussian distributions); (b) Switched Phase Modulation; and (c) Asynchronous Detection. We observe that the unconditional distributions are non-Gaussian, and also changes in the conditional marginal distributions in all three cases. Hence, according to Section \ref{subsec:theoryNG}, all the three non-Gaussian states have nonzero discord. }
\label{fig:NG1}
\end{figure*}

 1) \textit{Switched Noise Modulation} - The first non-Gaussian state is an equal statistical mixture of vacuum and a thermal state. The thermal state is produced by applying two independent Gaussian distributed noise signals to a phase and amplitude modulator. An external square wave modulation envelope at 12 kHz was then used to gate the two modulators. Square wave modulation turns the Gaussian modulation, on and off periodically. In this way the beam has either Gaussian modulation or no modulation at all. Since the square wave gating frequency is fast compare to the detection time, the net detected statistics seen will consist of an equal contribution from both the vacuum and the thermal state. Modulation and demodulation arrangement and the Wigner function of the produced state are shown schematically in Figure \ref{fig:ngschematic}(a).  The laser light with this non-Gaussian modulation then splits on a 50:50 beam splitter and each part is sent to a homodyne measurement station. To investigate the correlations between two subsystems, the same measurement procedure is performed as described in Section \ref{sec:expG}, and the results are presented in Figure \ref {fig:NG1} (a).
  
2) \textit{Switched Phase Modulation} - The second prepared non-Gaussian state is a mixture of vacuum and a coherent state. As depicted in Figure \ref{fig:ngschematic}(b), a sine wave modulation with frequency of 4 MHz is introduced to phase quadrature to create the coherent state. We then add a square wave modulation with frequency of 120 Hz to gate the sine modulation on and off. With this arrangement there is a sine modulation for half of the measurement time and no modulation for the other half. Signal is detected synchronously by using the same demodulation frequency as is used for modulation. Similar procedure is repeated to prepare a correlated bipartite state. In order to verify the presence of discord, the marginal distributions of Bob's state conditioned on two different sets of Alice's outcomes $x_{A}<-6$ and $x_{A}>-6$ are calculated and any possible correlation in conditional marginal distributions is investigated\footnote{As discussed in Section \ref{subsec:theoryNG}, in order to verify quantum discord in this class of non Gaussian states it is sufficient to calculate marginal distributions conditioned on any two sets of Alice's outcomes.}. The results are shown in Figure \ref {fig:NG1}(b).

3) \textit{Asynchronous Detection} - We prepare the third non-Gaussian state by using asynchronous detection. This is experimentally realised by choosing a demodulation signal different in the frequency by an small amount compared to the modulation signal. As displayed in Figure \ref{fig:ngschematic}(c), we drive the EOM by sine wave with frequency of 4 MHz and demodulate with frequency of 3.99MHz. The data collected is then digitally filtered to 3.9-4.1 MHz. The prepared state is a two peak probability distribution function along the X-quadrature as shown by Wigner function in Figure \ref{fig:ngschematic}(c) right. This is analogous to the  stroboscopic measurement of the quadrature of a harmonic oscillator. The marginal probability distribution of the prepared state and the conditional probability distributions are presented in Figure \ref {fig:NG1}(c).

As can be observed from Figure \ref{fig:NG1}, it is evident that the conditional probability distributions for all three non-Gaussian states are different from their unconditioned distributions. Neither their peaks nor the mean values of their distributions coincide, which by considering the preparation method, is a sufficient evidence of the presence of discord in the three non-Gaussian states. As the difference between two conditional marginal distributions is the criterion to verify quantum discord, in situations where the conditional distributions are very similar to each other, one can deploy $\chi^2$  test and calculate its probability function. Generally one rejects the \textit{null} hypothesis if the probability function is less than 0.05, which means two distributions are not the same. In our experiment, the calculated probability function is zero for all the states, indicating the two conditional distributions are completely different and the states are discordant.

\section{Conclusion}
\label{conclude}
We have introduced and experimentally demonstrated a simple and efficient method for verifying quantum discord in unknown bipartite Gaussian states. We have shown that by checking peak separation between the marginal distributions conditioned on two different homodyne measurements outcomes, the correlation of corresponding quadrature can be tested. With this technique, quantum discord can be verified by testing correlations between all four combinations of the amplitude and phase quadratures of two subsystems. By varying the modulation depth, we showed that our results are indeed consistent with the theoretical predictions within statistical errors. The robustness of our technique in small modulation depth permits one to detect nonzero discord even when the correlations are small. Moreover, we have discussed that our technique can be used for a certain class of non-Gaussian states. We applied our method to three different bipartite non-Gaussian states, which are prepared by subjecting statistical mixtures of coherent states to one port of beam splitter while the other port is in the vacuum state. Experimental results for all the non-Gaussian states show that the conditional marginal distributions are significantly different from the unconditional distributions, indicating nonzero quantum discord in each case. Our results show that with some prior knowledge about a quantum state, such as being Gaussian, or about the preparation stage quantum discord can be efficiently verified with a finite number of measurements.

\section*{ACKNOWLEDGEMENT}
We thank Mile Gu for fruitful discussion. The research is supported by the Australian Research Council (ARC) under the Centre of Excellence for Quantum Computation and Communication Technology (CE110001027).\\

\appendix
\section{Covariance matrix of the bipartite Gaussian state}
\label{appendix}
Covariance matrix of bipartite Gaussian state shown in Figure \ref{fig:schematic}(b) is 
\begin{equation}
\bm{\sigma}=
\begin{pmatrix}
15.96 & 0 &17.58 & 0 \\
0 & 14.37 & 0 & 13.55 \\
17.58 & 0& 22.62&0\\
0& 13.55&0&14.81
\end{pmatrix}
\;
\end{equation}
It can be seen from this covariance matrix that there are correlations between the quadratures of two subsystems ($\mathbf{C}\neq$ 0). Hence quantum discord is nonzero~\cite{rahimi-keshari2013}. It confirms our method that quantum discord is not zero when peaks of the conditional marginal distributions corresponding to two outcomes of homodyne measurements do not coincide.

\end{document}